\documentclass[12pt, preprint]{aastex} 
\def\degree{\ifmmode {^\circ}\else {$^\circ$}\fi}
\def\mum{\ifmmode {\rm \mu {\rm m}}\else $\rm \mu {\rm m}$\fi}
\def\arcsec{\ifmmode^{\prime \prime}\else $^{\prime \prime}$\fi}
\def\arcmin{\ifmmode^{\prime}\else $^{\prime}$\fi}
\def\kms{\ifmmode {\,\rm km \, s^{-1}}\else {$\, \rm km \, s^{-1}$}\fi }
\def\rsun{\ifmmode {\rm R_{\odot}}\else $\rm R_{\odot}$\fi}
\def\msun{\ifmmode {\rm M_{\odot}}\else $\rm M_{\odot}$\fi}
\def\lsun{\ifmmode {\rm L_{\odot}}\else $\rm L_{\odot}$\fi}

\def\ergcm2s{ergs~cm$^{-2}$~s$^{-1}$}
\def\cm2s{cm$^{-2}$~s$^{-1}$}
\def\redchi2{$\chi_\nu^2$}
\def\1543{4U~1543$-$47}
\def\nh{$N_{\rm H}$~}
\def\cm2{cm$^{-2}$} 
\shorttitle{AGN in EGS}
\shortauthors{Park et al.}

\begin{document}

\title{AEGIS: Radio and Mid-infrared Selection of Obscured AGN Candidates}
\author{S.~Q.~Park\altaffilmark{1},
        P.~Barmby\altaffilmark{2}, 
        G.~G.~Fazio\altaffilmark{1},
        K.~Nandra\altaffilmark{3},
        E.~S.~Laird\altaffilmark{3},
	A.~Georgakakis\altaffilmark{3},
        D.~Rosario\altaffilmark{4},
        S.~P.~Willner\altaffilmark{1},
        G.~H.~Rieke\altaffilmark{5},
	M.~L.~N.~Ashby\altaffilmark{1},
        R.~J.~Ivison\altaffilmark{6},
        A.~L.~Coil\altaffilmark{5,7},
        S.~Miyazaki\altaffilmark{8}}

\altaffiltext{1}{Harvard--Smithsonian Center for Astrophysics, 60 Garden Street,
 Cambridge, MA 02138, USA; spark@cfa.harvard.edu.}
\altaffiltext{2}{Department of Physics \& Astronomy, University of
 Western Ontario, London, ON N6A 3K7, Canada.}
\altaffiltext{3}{Astrophysics Group, Imperial College London, Blackett
 Laboratory, Prince Consort Road, London SW7 2AZ, United Kingdom.}
\altaffiltext{4}{UCO/Lick Obs., Univ. of California Santa Cruz, 1156
 High St., Santa Cruz, CA 95064, USA.}
\altaffiltext{5}{Steward Observatory, University of Arizona, 933
 N. Cherry Avenue, Tuscon, AZ 85721, USA.}
\altaffiltext{6}{UK Astron. Technology Ctr., Royal Obs., Blackford
 Hill, Edinburgh EH9 3HJ, United Kingdom.}
\altaffiltext{7}{Hubble Fellow.}
\altaffiltext{8}{Subaru Telescope, National Astronomical Observatory
 of Japan, 650 North A`ohoku Place, Hilo, HI 96720, USA.}

\begin{abstract}
The application of multi-wavelength selection techniques is crucial
for discovering a complete and unbiased set of Active Galactic Nuclei
(AGNs).  Here, we select a sample of 72 AGN candidates in the Extended
Groth Strip (EGS) using deep radio and mid-infrared data from the Very
Large Array (VLA) and the \emph{Spitzer Space Telescope}, and analyze
their properties across other wavelengths.  Only 30\% of these sources
are detected in deep 200 ks \emph{Chandra X-ray Observatory}
pointings.  The X-ray detected sources demonstrate moderate
obscuration with column densities of \nh $\gtrsim 10^{22}$ \cm2.  A
stacked image of sources undetected by \emph{Chandra}  shows low levels of
X-ray activity, suggesting they may be faint or obscured AGNs.  Less
than 40\% of our sample are selected as AGNs with optical broad lines,
mid-infrared power laws, or X-ray detections.  Thus, if our
candidates are indeed AGNs, the radio/mid-infrared selection criteria
we use provide a powerful tool for identifying sources missed by other
surveys.
\end{abstract}


\maketitle
%----------------------------------------------------------------------

\section{INTRODUCTION}
Active Galactic Nuclei (AGNs) are sources of tremendous energy that
provide clues to the accretion history of the Universe.  Defining a
complete and reliable sample of AGNs is of great importance in our
understanding of how black holes grow over cosmic time as well as the
physical processes behind these powerful objects.  X-ray emission is
an efficient means of detecting energetic activity in distant galactic
cores and is often used to find AGN samples \citep{mus04}.  However,
this approach is less successful at finding sources hidden by extreme
columns of gas and dust, and models of the observed hard X-ray
background require larger numbers of AGNs than are currently observed.
\citet{wor05} found that only 50\% of the X-ray background is resolved
at $> 8$ keV, and the missing component of the background is
consistent in spectral shape with a population of obscured AGNs at
redshifts $z \sim 0.5-1.5$ with \nh $ \sim 10^{23}-10^{24}$ \cm2.
These synthesis models predict that at 30 keV, the peak of the hard
X-ray spectrum, the X-ray background is dominated by heavily obscured
sources and obscured AGNs outnumber unobscured AGNs by a ratio of
three or four to one \citep{ued03, gil07, tre05, com01}.  Thus, the
bulk of the accretion in the Universe is taking place in obscured
environments that are likely to be missed by X-ray selection methods,
highlighting the need for a multi-wavelength approach to build an
unbiased census of AGNs.

Mid-infrared and radio surveys provide a way of searching for these
elusive obscured nuclei, as these wavelengths are much less affected
by extinction and can penetrate through the dusty atmospheres
surrounding the AGN.  X-ray obscured AGNs are often bright in the
mid-IR as the dusty environment surrounding the accretion disk
reprocesses incident X-rays from the central engine and re-radiates
these photons at infrared wavelengths.  Efforts to select AGN
candidates with infrared data have mainly concentrated on using mid-IR
color criteria \citep[e.g.,][]{lac04,ste05} or power-law selection
techniques \citep[e.g.,][]{alo06,bar06}.  Recently, \citet{dad07} and
\citet{fio07} presented evidence of finding Compton thick AGN
candidates (\nh $> 10^{24}$) by searching for sources with large
mid-IR excess.

Successful attempts have also been made to find obscured AGNs with
combined analyses of mid-IR and radio data sets. \citet{don05} found a
minimum 4:1 ratio of obscured (Compton thin and Compton thick) to
unobscured AGNs in the \emph{Chandra} Deep Field-North (CDF-N) by
analyzing sources with radio excess.  \citet{mar05} found a 2$-$3:1
ratio using a combination of mid-IR, near-IR, and radio flux density
criteria in the \emph{Spitzer} First Look Survey (FLS) and the Subaru
XMM-Newton Deep Field (SXDF).  Both results are within model
predictions of the obscured AGN ratio.  Here, we combine and compare
the procedures introduced by these two radio/mid-IR selection
techniques to find a sample of 72 AGN candidates in the Extended Groth
Strip (EGS) and use multi-wavelength data to cross-identify their
active status.  In particular, understanding the X-ray properties of
the candidate AGNs is important to control any residual contamination
by non-AGNs as well as to study their obscuration and thus put more
constraints on the X-ray background.  Throughout this paper, we adopt
$\Lambda$ cold dark matter cosmology with $H_0 = 70$ km s$^{-1}$
Mpc$^{-1}$, $\Omega_M = 0.3$, and $\Omega_{\Lambda} = 0.7$.

\section{OBSERVATIONS AND DATA REDUCTION}

The Extended Groth Strip (EGS) is an approximately 2\degree~by
15\arcmin~region centered on $\alpha = 14^h17^m$ and $\delta =
+52\degree30\arcmin$.  The region's high ecliptic and Galactic
latitudes make it ideal for studying extragalactic sources with little
Galactic contamination, attracting deep coverage of the field in all
wavelengths from radio to X-ray.  It is a unique field in its
combination of size and depth: it is shallower but larger than the
combined GOODS fields, and smaller but deeper than the COSMOS field at
most wavelengths.  Thus, the intermediate position of the EGS in both
depth and area makes it an ideal dataset for combining the AGN
selection techniques previously tested in different regimes.
\citet{dav07} give a detailed description of the region and the
All-Wavelength Extended Groth Strip International Survey (AEGIS) data
sets.

The radio data were observed at 1.4~GHz using the Very Large Array
(VLA) with $\sim$18 hr of data per sky position and a resulting source
sensitivity limit of 50 $\mu $Jy \citep{ivi07}.  The data have a noise
level of 10 $\mu $Jy beam$^{-1}$ and a FWHM of $\sim 3.8$\arcsec.
Sources were defined by searching signal-to-noise ratio images using
the SAD detection algorithm; the technique is outlined in
\citet{big06}.  The infrared properties were found using instruments
aboard the \emph{Spitzer Space Telescope} \citep{wer04}, with 1.2 ks
observations from the Multiband Imaging Photometer
\citep[MIPS,][]{rie04} and 9.1 ks exposures from the Infrared Array
Camera \citep[IRAC,][]{faz04}.  The \emph{Spitzer} data were processed
with the MIPS Instrument Team Data Analysis Tool and the Spitzer
Science Center IRAC pipeline, and sources were identified using {\sc
daophot}.  Further details for the MIPS data are provided by
\citet{eng07} and for the IRAC data by Barmby et al.~(2008).  The
5$\sigma$ limiting point-source flux densities of the infrared data
are 83 $\mu$Jy for the 24~\micron\ MIPS band and 2 $\mu $Jy for the
3.6~\micron\ IRAC band.  Their FWHMs are 5.9\arcsec and 1.8\arcsec at
24~\micron\ and 3.6~\micron, respectively.

The radio and infrared catalogs were matched to each other using a
2\farcs5 search radius.  The overlapping
VLA and MIPS images result in a field of view of $\sim 0.35$ deg$^2$,
with roughly 630 and 9,800 sources at 1.4~GHz and 24~\micron,
respectively. In this shared area, there are 482 sources with both
1.4~GHz and 24~\micron\ detections; these sources form the parent
sample for our selection.  This paper uses two radio/mid-IR selection
techniques, one of which also requires use of the 3.6~\micron\ data.
Due to a small offset in angle between the IRAC and MIPS images, the
IRAC field does not extend as far north as the MIPS and radio fields.
Thus, only 325 of the sources detected in both 1.4~GHz and 24~\micron\
are within the IRAC field of view; 321 of these have 3.6~\micron\
counterparts.  (The remaining four sources are visible on the IRAC
image but were not extracted by {\sc daophot}.)

After selecting our candidates, we analyzed their properties using 200
ks X-ray data from the ACIS-I instrument on the \emph{Chandra X-ray
Observatory}.  The X-ray image is aligned with the IRAC field and thus
does not completely cover the VLA/MIPS shared area.  Data reduction
methods are described by Nandra et al.~(2007 in prep).  We also
utilize \emph{R} band photometry covering the entire field to a
5$\sigma$ limiting AB magnitude of $\sim$26.5 from the Suprime camera
on the Subaru telescope (S.~Miyazaki 2005, private communication).
Supplemental $R$ band photometry from the Canada-France-Hawaii
Telescope (limiting AB magnitude = 24.2 at 8$\sigma$; \citet{coi04})
was used in a few cases where the Subaru data lacked coverage.
Finally, we use redshifts obtained by the DEEP2 survey, available in
two-thirds of the VLA/MIPS shared field from the Keck DEIMOS
spectrometer \citep{dav03}.  The median redshift for the survey is $z
\sim 0.7$; redshift measurements for the candidate
sources ranged from $z \sim 0.9-1.4$.  Because the optical fields are quite
crowded, sources with multiple potential counterparts within 2\farcs5
were checked by eye.  Two optical counterparts were thrown out because
the match could not be determined unambiguously.

\section{AGN SELECTION TECHNIQUES}
Two different samples of radio/mid-IR AGN candidates were selected
using two separate sets of criteria, as illustrated in Figure 1.  The
first method, which we refer to as ``$q$-selection'', relies on the
radio/infrared relation for star-forming galaxies and AGNs
\citep{con92}.  Recent findings have suggested that the relation may
break down at high redshift \citep{kov06, vla07}, but \citet{sey07}
show that the correlation holds to at least $z \sim 3$.  Their study
compares observed 1.4 GHz with observed 450 and 850 \micron\ flux
densities, which at the redshifts of interest ($z > 1.5$) sample rest
wavelengths in the radio near 5 GHz and in the far infrared near 150
and 300 \micron.  \citet{app04} defined $q$ =
log(S$_{24~\mum}$/S$_{1.4~\rm GHz})$ and showed that $q = 0.84 \pm
0.28$ for non-K-corrected normal star-forming galaxies.  \citet{don05}
suggested that sources with strong radio emission relative to their
24~\micron\ flux densities are likely to be dominated by AGNs rather
than star formation and analyzed a sample of candidate AGNs by
selecting sources with $q < 0$. Of the 482 sources detected at both
1.4~GHz and 24 \micron\ in our field, 51 have $q < 0$.  We discarded
three sources that had flux errors too large to yield convincing
ratios of $q < 0$.  A fourth source was discarded because the radio
detection had only a marginal 24~\micron\ counterpart ($d=2\farcs47$)
and there were too many possible optical counterparts to be confident
in the match.  The $q$-selection only requires the use of the VLA and
MIPS parent sample.  Unlike the sample of \citet{don05}, we require
all our candidates to have a detection in the 24~\micron\ band in
order to better compare with the second selection technique of this
paper.  Because of the offset between the MIPS and IRAC images, 35\%
of these sources lie outside the IRAC field, and 20\% are outside the
\emph{Chandra} field.

The $q$-selected sample should include few if any starburst
contaminants: adopting $\sigma = 0.28$ at $z = 0$ from \citet{app04},
a threshold of $q < 0$ selects sources 3.4 standard deviations below
the average $q$ value expected for starbursts.  This difference may
even be a conservative estimate, as \citet{boy07} find a higher $q =
1.39 \pm 0.02$ for starbursts using stacked data in the \emph{Spitzer}
Wide Field Survey (SWIRE).  \citet{don05} and \citet{sey07} show that
at higher redshifts, the difference in $q$ should remain constant or
increase when comparing LINERs, Seyfert 2s and spiral galaxies to
starburst systems.  The $q$ values of ULIRGs will fall closer to our
selection limit with increasing redshift, though their values should
remain above our cut of $q = 0$ out to $z \gtrsim 2.5$.  At such high
redshifts, even our faintest candidates would have radio luminosities
too high to be consistent with starburst systems, and we do not expect
our sample to be contaminated with such non-AGN sources.

The second selection method, hereafter referred to as the ``flux-cut''
method, employs the 3.6~\micron\ data set in addition to the
24~\micron\ and 1.4~GHz parent population to identify an AGN-rich
sample.  The selection is designed to find obscured AGNs near $z \sim
2$ and is thus a good complement to the lower-$z$ $q$-selection
technique.  The redshift calculation comes from assuming the
3.6~\micron\ emission will be dominated by the old stellar population
and forming a 3.6~\micron-$z$ relation \citep{mar05}, analogous to the
$K$-$z$ relation for radio galaxies \citep{wil03}.  The relation
assumes the galaxies have an elliptical host galaxy, which would not
be the case for potential starburst contaminants.  However, comparing
the relation with local galaxy templates revealed that the correlation
(and thus, the redshift assumption) remained valid even for reddened
ULIRG templates.  The flux-cut method is outlined in detail by
\citet{mar05, mar07}.  We adopted their flux limits:

\indent$(i)~~	\rm S_{24~\mum} > 300~\mu $Jy \\
\indent$(ii)~~	100~\mu $Jy$ \le \rm S_{1.4~\rm GHz} \le 2~$mJy \\
\indent$(iii)~~	\rm S_{3.6~\mum} \le 45~\mu $Jy \\

The 24~\micron\ selection was chosen to find a flux density limited
sample of active galaxies; a flux density of 300 $\mu$Jy at $z=2$
corresponds to quasars of luminosity $L \ge 0.2 L^*_{\rm quasar}$ at
rest-frame 8~\micron.  At these wavelengths, dust extinction is
negligible except in extreme cases.  In AGN unification schemes, a
dusty torus surrounds the optically and X-ray bright accretion disk.
Dust particles absorb and re-emit these photons, and a torus viewed
edge-on would therefore be optically/X-ray obscured but \emph{bright}
in the mid-IR.  \citet{wee06} and \citet{bra06} show that the AGN
fraction tends to increase as $24~\mum$ flux density increases.
Accepting only higher flux density sources should therefore improve
our probability of finding AGNs.

At $z \sim 2$, the MIPS criterion alone may select star-forming
systems with strong PAH features.  The minimum radio flux density cut
is designed to eliminate these starburst systems.  Using the lower
limit of 100~$\mu$Jy from \citet{mar07} rather than the 350~$\mu$Jy
limit of \citet{mar05} increases our sample size while still being
high enough to avoid nearly all starburst contaminants.  Using the
equations for star-forming rates from \citet{yun01} and \citet{wu05},
if we assume that the 24~\micron\ emission of 300~$\mu$Jy arises
solely from star-forming systems at $z \sim 2$, we would expect a
radio flux density that is still an order of magnitude lower than 100
$\mu$Jy.  The upper limit of the radio criterion removes extreme
radio-loud sources whose potential jets may complicate the analysis;
only one potential flux-cut candidate is omitted by this upper limit
(see Figure 1).  From the parent sample of 482 sources, 184 fit these
MIPS and radio flux criteria.

The 3.6~\micron\ criterion weeds out unobscured AGNs and corresponds to
a rest frame wavelength of 1.2~\micron\ at $z=2$.  Dust extinction
makes the near-IR emission of obscured AGNs much fainter than that of
unobscured sources.  Requiring high 24~\micron\ flux densities and low
3.6~\micron\ flux densities focuses the sample on obscured types.  Out
of the 184 sources that fit the 24~\micron\ and 1.4~GHz criteria, 115
are within the IRAC field.  Of those, the 3.6~\micron\ criterion
yields a final sample size of 28 flux-cut sources.  We compared the
radio/24~\micron\ ratios of these sources with the values expected
from local ULIRG templates and analyzed their SEDs and
multi-wavelength properties.  The results suggested only 2-3 potential
starburst candidates among the sources; these few interlopers should
not alter our results.

We combined both of these techniques to arrive at our final sample.
Because we are likely probing different redshift regimes and
obscuration levels between the two techniques, only three sources
overlap between the 47 $q$-selected sources and the 28 flux-cut
sources.  Our final sample thus contains 72 radio/mid-IR AGN
candidates (see Table 1).

\section{MULTI-WAVELENGTH PROPERTIES}

\subsection{X-Ray Properties}

Thirty-seven of the 47 $q$-selected sources are within the
\emph{Chandra} field; 15 (41\%) have X-ray detections with Poisson
false probability $< 10^{-3}$ (corresponding to slightly above
3$\sigma$ in Gaussian statistics) in at least one X-ray band.  Of the
28 flux-cut sources, 26 lie within the X-ray image, and four have
X-ray counterparts (15\%).  (The detection rates are only 32\% and
12\% for the $q$-selection and flux-cut methods in the higher
significance AEGIS-X catalogs (Nandra et al.~2007, in prep), which
require a more stringent Poisson cut of $< 4 \times 10^{-6}$.)  In
comparison, \citet{don05} employed the $q$-selection method in the
CDF-N, where VLA and MIPS observations are similar in depth to ours
but \emph{Chandra} X-ray exposures exceed 1 Ms.  They found an X-ray
counterpart detection rate of 40\% at Poisson probability $< 3 \times
10^{-7}$ (5$\sigma$), and the rate rises to nearly 80\% for
probability $< 2 \times 10^{-2}$ (2$\sigma$) detections.
\citet{mar07} detected $\gtrsim 5$\% of their flux-cut-selected
candidates with a flux limit F$_{2-12~\rm keV} = 3 \times 10^{-15}$
\ergcm2s in the shallower SXDF with XMM exposures of $< 100$ ks.  They
were able to increase the detection fraction to 17\% by fitting SEDs
to filter out contaminants, a procedure we do not adopt here.

As expected, deeper surveys are required to detect X-ray counterparts
to greater completion.  Although the \emph{Chandra} data in the
EGS comprise the third deepest \emph{Chandra} survey to date, 70\% of
our sources lack counterparts at the 200 ks depth.  This may be due in
part to the relative depth of the \emph{Spitzer} data versus the
\emph{Chandra} field (see Figure 2).  In the EGS, over 50\% of the
X-ray sources in the field have MIPS counterparts, but only 5\% of the
MIPS sources have X-ray counterparts.  However, some of the undetected
sources have upper limits that are greater than the fluxes of the
actual detected sources, and often lie on relatively shallow
(off-axis) areas of the X-ray map. These sources would likely have
been detected with deeper data or on-axis pointings.

X-ray hardness ratios are often used to study AGN type, as harder
spectra generally indicate more obscuration.  All four X-ray sources
in the flux-cut method were detected only in the soft band, whereas 11
of the 15 sources detected with the $q$-selection method had both hard
and soft band detections.  Two other $q$-selected sources were
detected only in the soft band, one was detected only in the hard
band, and the final source was barely detected in the full band but
had neither hard nor soft band detections at our limits. Excluding the
source with a full band only detection, the $q$-selected sources have
a mean hardness ratio of $-0.19$ (HR = H$-$S / H$+$S, where H = $2-7$
keV counts and S = $0.5-2$ keV counts).

As measured by hardness ratios, one-third of the X-ray detected
$q$-selected sources are obscured according to the division chosen by
\citet{szo04}, who classify obscured AGNs as those having HR $> -0.2$
in the \emph{Chandra} Deep Field South.  Our obscured fraction would
rise if we include $q$-selected sources not detected in X-ray.
Assuming all X-ray undetected sources are obscured would yield a
maximum obscuration fraction of 73\%.  At $z=1$, we derive X-ray
luminosities of $10^{42}$ to $10^{44}$ ergs/s from the observed fluxes
of the detected sources.  At this redshift, an obscuration fraction of
73\% would be within the limits of \citet{szo04}, who found
~45\%$-$75\% of their sources to be obscured in the same X-ray
luminosity range.

Overall, the combined AGN candidates we find tend to have
\emph{softer} hardness ratios than the general sample of X-ray sources
in the AEGIS-X catalog.  Including the four soft flux-cut sources with
the $q$-selected sources, the mean hardness ratio falls to $-0.37$,
while the total X-ray population has a mean HR = $-0.17$.  The general
population of sources with both X-ray and radio detections are even
harder with a mean HR $= -0.12$.  However, a lack of hard band
detection is not necessarily an indication that the source has a soft
spectrum, especially for faint sources at higher redshifts
\citep{lai06}.  Also, the $q$-selected AGNs generally have fluxes
comparable to the entire X-ray sample, with an average flux
F$_{2-10~\rm keV} = 8.3 \times 10^{-15}$ \ergcm2s in the hard band and
F$_{0.5-2~\rm keV} = 2.7 \times 10^{-15}$ \ergcm2s in the soft band.
On the other hand, the flux-cut sources are fainter by a factor of 10
in the soft band (and not detected at all in the hard band).  This
weaker emission, coupled with the expectation that these flux-cut
sources should be near $z \sim 2$, suggests that the sources might be
too faint for hard band detections, and spectral fitting is necessary
to determine their true spectral slopes.

Two of the X-ray counterparts with soft and hard band detections have
spectroscopic redshifts; both of these are obscured with column
densities $\gtrsim 10^{22}$ \cm2, assuming a power-law
model\footnote[9]{Column densities were estimated using PIMMS
\citep[Portable, Interactive Multi-Mission Simulator;][]{muk93} v3.9b
to convert between X-ray source fluxes and count rates.} with
intrinsic $\Gamma = 1.9$ \citep[e.g.][]{nan97}.  Nine X-ray
counterparts had both soft and hard band detections but no optical
spectral information.  We estimate column densities for these assuming
various redshifts.  At $z = 0$, four of the nine sources would be
obscured with \nh $> 10^{22}$ \cm2.  At $z = 1$, seven sources would
be obscured, and at $z = 2$, eight sources would show obscuration,
though none Compton-thick.  The ninth source, which does not show
obscuration, is a very bright, nearby galaxy with a hardness ratio of
$-0.46$.  This source is likely a normal galaxy and is further
discussed in Section 4.2.

\subsubsection{X-Ray Stacking}

For the sources without X-ray detections we use a stacking analysis to
search for traces of X-ray emission \citep[see][]{geo08}.  Removing
sources outside the \emph{Chandra} field and sources too close to a
bright, potentially contaminating neighbor yielded 39 stackable
sources.  Stacking the counts from the 39 undetected sources yielded
significant detections ($> 3\sigma$) in both the hard and soft bands.
The mean hard band flux of the stacked galaxies is $(1.55 \pm 0.50)
\times 10^{-16} $ \ergcm2s and the mean soft band flux is $(3.73 \pm
0.87) \times 10^{-17}$ \ergcm2s.  These fluxes are near the detection
limit of the 2Ms CDF-N, and deeper observations in the EGS would
likely have detected more of these stacked sources. However, the
average fluxes indicate that as many as half of these sources would
not have been detected at the catalog limits even in the deepest
existing X-ray survey. The work of Donley et al. (2005, 2007) suggests
that most would be weakly detectable ($\ge 2 \sigma$) in the deepest
surveys, with a minority well below even these limits.

The stacked source has a hardness ratio of $-0.11 \pm 0.20$
corresponding to an effective X-ray spectral slope $\Gamma$ = 1.2,
which is slightly harder than the $\Gamma = 1.4$ observed for the
X-ray background.  Assuming a redshift $z=1$, the observed hardness
ratio corresponds to a moderately obscured column density of $\sim 4
\times 10^{22}$ \cm2.

The distribution of detected counts in the stacked signal indicates
that the emission is not dominated by a small sample of sources, but
is representative of the overall undetected population as a
whole. These results are for the combined sample; repeating the
analysis for the undetected flux-cut and $q$-selected sources
separately yields roughly similar spectra and fluxes, though the
flux-cut sample is slightly harder with HR = $-0.05$ (corresponding to
an effective $\Gamma = 1.1$), as opposed to HR = $-0.15$
(corresponding to an effective $\Gamma = 1.3$) for the $q$-selected
sample.

The results of the stacking analysis suggest that the undetected
sources may be faint or moderately obscured AGNs.  However, the
spectral slopes are too uncertain to accurately determine
the fraction of Compton thick sources in our sample.

\subsection{Optical Properties}
Eight of the 72 candidates have associated DEEP2 spectra, six from the
$q$-selected sample and two from the flux-cut sample.  Two
$q$-selected sources (33\%) and one flux-cut source (50\%) show
optical signatures of AGNs, with broad emission lines and/or strong
absorption features indicative of outflows.  All three of these
optically identified AGNs also have X-ray counterparts.  One optical
source fits the spectrum of a normal massive elliptical galaxy.  The
rest show either no optical spectral features, or features resembling
star-forming systems.  If these objects are AGNs, our selection
criteria find sources that mostly have no optical AGN signatures and
may be part of the obscured population.

Four $q$-selected sources have spectroscopic redshifts ranging from 0.89
to 1.41 with a mean of $z \sim 1$. Two sources from the flux-cut
sample have spectroscopic redshifts, both near $z \sim 1$.
Assuming the 3.6 \micron\ emission is largely dominated by the old
stellar population rather than AGN light, we can roughly estimate the
photometric redshift of the remaining sources using the $K-z$ relation
for radio galaxies \citep{wil03}.  With our assumption, the 3.6
\micron\ flux densities are upper limits to those at $K$, so the flux
densities for the IRAC-detected sources imply redshifts $z \ge 0.5$.
If there is an AGN contribution at 3.6 \micron\ (e.g., the power law
sources discussed below), then the estimated redshifts become lower
limits.  If we assume similar host galaxies for the flux-cut sample,
the selection limit S$_{3.6~\mum} < 45 \mu$Jy yields expected
redshifts $z \gtrsim 1$.  \citet{mar05} adapt the $K-z$ relation to
define an $S_{3.6~\mum} - z$ correlation which predicts $z \gtrsim
1.4$ for the flux-cut sample.

The hard X-ray/optical \emph{R} band emission ratio (X/O; see Figure
3) is a diagnostic widely used to separate AGNs from starburst
galaxies \citep{mac88, ale01}.  Normal galaxies tend to have very weak
hard X-ray emission and thus very small X/O ratios.  However, AGNs are
able to fuel energetic X-rays and are expected to have X-ray to
optical ratios of $\sim 0.1-10$.  An X/O ratio of $0.01-0.1$ is
indicative of lower luminosity AGNs or galaxies dominated by
star-formation.  Of the 12 candidate sources with hard X-ray and
\emph{R} band detections in our sample, 10 have X/O $> 0.1$, and 8
have X/O $> 1$.  Two sources in our sample have extreme X-ray to
optical emission ratios $> 10$.  In the general population of sources
with X-ray and optical detections, we found that only 55\% of sources
had X/O $> 1$.  However, our sample sizes are too small to draw
statistically significant conclusions between our candidates and the
general population.  Figure 3 shows the expected
X/O ratios from the stacked signals; their positions are consistent
with the sources being AGNs rather than starbursts.

One $q$-selected source, mentioned in Section 4.1, has a very small
X/O ratio and is extremely bright in the optical with $R = 16.1$.  The
hard X-ray flux is F$_{2-10~\rm keV} = 1.0 \times 10^{-14}$ \ergcm2s.
This source is the only X-ray source that does not show obscuration.
It is not plotted in Figure 3 as it is three orders of magnitude
brighter in $R$ than the brightest source in the figure.  The X/O
ratio of this source and the soft hardness ratio suggest that the
X-ray emission may be dominated by X-ray binaries and hot gas in a
normal galaxy rather than an active nucleus \citep{hor03}.

\subsection{Infrared Properties}
Figure 2 plots the hard X-ray flux of the X-ray detected sources
against the 24~\micron\ flux density.  Half of the detected sources
lie in the designated AGN region for local galaxies as determined
observationally by \cite{pic82} while the other half are in the
transition region. The stacked signal is also in the transition
region, though close to the local starbursts. Though we have not
corrected our sources for redshift, \citet{ale01pic} show that the
observed hard X-ray to mid-IR flux ratio will minimally change with
increasing redshift for sources with low column densities and will
increase more noticeably for obscured AGNs.

Many AGNs can also be described by their characteristic negative
power-law spectral energy distributions (SEDs) which extend from
mid-IR to ultraviolet wavelengths: $f_\nu \propto \nu^\alpha$, with
$\alpha < -0.5$ \citep[][see also Alonso-Herrero et al.~2004 and
Barmby et al.~2006]{elv94}.  This emission can arise from multiple
source components near the dusty nucleus that broaden the SED rather
than from a single power-law source \citep[e.g.][]{rie81}.  In the
IRAC bands, stellar-dominated objects will either exhibit positive
(blue) power-law behavior with $\alpha \sim +2$ corresponding to the
Rayleigh-Jeans tail of the blackbody spectrum or else not fit a power
law at all due to their 8~\micron\ PAH emission or 1.6~\micron\
stellar bump.  A number of $q$-selected sources lie outside the IRAC
observed field or lack detections in all four IRAC bands, leaving a
sample of 53 out of 72 candidates for which we could study IRAC
power-law properties.  Ten of these (19\%) are well fit to a power law
between 3.6~\micron\ and 8.0 \micron\ with $\alpha < -0.5$.  Examples
of such fits can be found in \citet{alo06}, \citet{bar06}, and
\citet{don07}.  The
detection rate is approximately the same for both selection methods,
and is comparable to the fraction of X-ray sources with similar
power-law slopes in the EGS.  \citet{don07} show that the majority of
power-law galaxies in the CDF-N lie at high redshifts.  The low
fraction of power-law detection may be due to the fact that most of
our sample are q-selected galaxies, which are predicted to be at
low-moderate redshifts.

In a random sample of 53 IRAC sources, we would expect $\sim$2 sources
to exhibit negative IR power-law behavior based on the numbers found
for the entire IRAC population.  The Poisson probability of finding 10
or more such power-law sources, as we have done, is only $10^{-4}$.
Of the general sample of 321 sources with radio, MIPS and IRAC
detections, 23 have IRAC power-laws.  Thus, the overlap between our
sample and the mid-IR power-law sample is substantial, and suggests
that both methods are effective at finding AGNs, but the populations
found are distinct enough to make both methods necessary.

\section{SUMMARY AND DISCUSSION}
We have selected 72 AGN candidates by combining and comparing two sets
of radio and mid-IR selection criteria.  The $q$-selection method
finds 47 sources with 1.4~GHz flux densities far in excess of their
24~\micron\ flux densities, relative to star-forming galaxies.  The
flux-cut method selects 28 sources based on their 3.6~\micron,
24~\micron, and 1.4~GHz flux densities.  Only three sources overlap in
the two methods, since they are designed to probe different redshifts
and obscuration levels.  This selection is independent of typical
selection techniques such as X-ray or optical identifications, and
thus we can test the completeness of current AGN samples by finding
AGNs missed at other wavelengths.

AGNs are often defined via their mid-IR power-law behavior, optical
broad lines, and X-ray detections.  Nineteen sources (30\%) in our
sample have X-ray detections with Poisson false probabilities $<
10^{-3}$ at the \emph{Chandra} 200 ks depth.
Column density estimates of the X-ray detected sources indicate that
even assuming $z = 0$, half of our sources would be obscured with \nh
$> 10^{22}$ \cm2.  The obscuration fraction will increase with
redshift.  If we assume our sources are at $z = 1$, which is the mean
value determined from the few sources with spectroscopic data, nearly
all of the X-ray detected sources are obscured. Among the X-ray
undetected sample, stacking techniques demonstrate low but significant
levels of X-ray activity above the background emission.  The stacked
source has $\Gamma = 1.2$, which is slightly flatter (harder) than the general
X-ray background.  The relatively high X-ray detection fraction and
the properties of the stacked source indicate that we are finding a
population of moderately obscured, though not Compton-thick,
previously undetected AGNs.  The rate of X-ray confirmation of AGN
status is a function of the depth of the observations, and future work
in this area will benefit from recently awarded additional
\emph{Chandra} pointings in the field.

The radio/mid-IR selected sample also finds more sources separately
defined as AGNs via their optical or infrared properties than a random
selection from the parent sample.  However, our sample is also
distinct from AGNs found using these other selection techniques: less
than 20\% exhibit the IRAC power-law behavior expected of luminous AGNs
\citep[e.g.,][]{don07}, and 38\% show AGN signatures in their optical
spectra.  Table 1 summarizes the numbers of AGNs identified by each
technique separately for the flux-cut and $q$-selected samples.
Accounting for the overlap in multi-wavelength AGN identification for
some sources, 60\% of our candidates have not been identified as AGNs
via optical spectra, IRAC power-laws, or X-ray detections. The
selection criteria used here thus contribute to a more complete census
of AGNs, and a variety of selection techniques, requiring study at all
wavelengths, are required in order to find the least biased samples of
AGNs.

Compared to our X-ray detection rate of 30\%, \citet{don05} detected
80\% (40\%) of the candidates in their $q$-selected sample at
$2\sigma$ ($5\sigma$) in the deeper CDF-N, while \citet{mar07}
detected $\gtrsim$ 5\% in the shallower SXDF with a flux limit of
F$_{2-12~\rm keV} = 3 \times 10^{-15}$ \ergcm2s.  
In general, the flux-cut method finds a greater proportion of AGNs not
identified via other wavelengths than the $q$-selected sources.  This
may be due to a greater fraction of obscured sources in the sample, as
indicated by their stacked signal being slightly harder than the
$q$-selected stacked signal.  Also, the flux-cut sample is much
fainter at X-ray and optical wavelengths and AGN activity may be too
faint to detect in X-rays (either intrinsically, or due to larger
redshifts) or hidden by host galaxy light at optical wavelengths.
However, it is also possible that the flux-cut sample simply has a
higher rate of contamination from starforming systems, and thus a
smaller fraction of confirmed AGNs.

Both of the selection techniques we apply find only sources that are
either moderately radio-loud or occupy the bright end of the
radio-quiet population \citep{ive02}.  Radio-loud sources are expected
to make up only $\sim 10-20$\% of the total AGN population
\citep{kel89, urr95, ive02}, so 80-90\% of AGNs are undetected with
these techniques.  Sixty per cent of our candidates remained
undetected using other AGN identification techniques.  If we assume
that under the unified model of AGNs, the central engine behaves the
same for both radio-loud and radio-quiet sources, we can estimate that
the same fraction of radio quiet sources are undetected by these other
methods.  This implies that half of the overall AGN population has yet
to be found, and novel techniques must be developed in order to obtain
a complete sample of AGNs.

\acknowledgements We are grateful to the referee for comments that
helped improve this paper.
This work is based in part on observations made with
the Spitzer Space Telescope, which is operated by the Jet Propulsion
Laboratory, California Institute of Technology under a contract with
NASA. Support for this work was provided by NASA.  This work is based
in part on data collected at Subaru Telescope, which is operated by
the National Astronomical Observatory of Japan.  ESL acknowledges
support from STFC. AG acknowledges funding from the Marie-Curie
Fellowship grant MEIF-CT-2005-025108.  ALC is supported by NASA
through Hubble Fellowship grant HF-01182.01-A, awarded by the Space
Telescope Science Institute, which is operated by the Association of
Universities for Research in Astronomy, Inc., for NASA, under contract
NAS 5-26555.

Facilities: \facility{CFHT}, \facility{CXO}, \facility{Keck:II},
\facility{Spitzer (IRAC, MIPS)}, \facility{Subaru (Suprime)}, \facility{VLA}

\bibliographystyle{apj}

\begin{thebibliography}{5}

\bibitem[Alexander et al.(2001)]{ale01pic} Alexander, D.~M., et al.\ 2001, \apj, 554, 18 
\bibitem[Alexander et al.(2001)]{ale01} Alexander, D.~M., Brandt, W.~N., Hornschemeier, A.~E., Garmire, G.~P., Schneider, D.~P., Bauer, F.~E., \& Griffiths, R.~E.\ 2001, \aj, 122, 2156 
\bibitem[Alonso-Herrero et al.(2004)]{alo04} Alonso-Herrero, A., et al.\ 2004, \apjs, 154, 155 
\bibitem[Alonso-Herrero et al.(2006)]{alo06} Alonso-Herrero, 
A., et al.\ 2006, \apj, 640, 167 \bibitem[Appleton et al.(2004)]{app04} Appleton, P.~N., et al.\ 2004, \apjs, 154, 147 
\bibitem[Barmby et al.(2006)]{bar06} Barmby, P., et al.\ 2006, \apj, 642, 126 
\bibitem[Barmby et al.(2007)]{bar07} Barmby, P., et al.\ 2008, \apjs,
  submitted
\bibitem[Biggs \& Ivison(2006)]{big06} Biggs, A.~D., \& Ivison, R.~J.\ 2006, \mnras, 371, 963 
\bibitem[Brand et al.(2006)]{bra06} Brand, K., et al.\ 2006, \apj, 644, 143 
\bibitem[Boyle et al.(2007)]{boy07} Boyle, B.~J., Cornwell, T.~J.,
  Middelberg, E., Norris, R.~P., Appleton, P.~N., \& Smail, I.\ 2007,
  \mnras, 376, 1182
\bibitem[Coil et al.(2004)]{coi04} Coil, A.~L., Newman, 
J.~A., Kaiser, N., Davis, M., Ma, C.-P., Kocevski, D.~D., \& Koo, D.~C.\ 
2004, \apj, 617, 765\bibitem[Comastri et al.(2001)]{com01} Comastri, A., Fiore, F., Vignali, C., Matt, G., Perola, G.~C., \& La Franca, F.\ 2001, \mnras, 327, 781 
\bibitem[Condon(1992)]{con92} Condon, J.~J.\ 1992, \araa, 30, 575 
\bibitem[Daddi et al.(2007)]{dad07} Daddi, E., et al.\ 2007, ArXiv e-prints, 705, arXiv:0705.2832 
\bibitem[Davis et al.(2003)]{dav03} Davis, M., et al.\ 2003, \procspie, 4834, 161 
\bibitem[Davis et al.(2007)]{dav07} Davis, M., et al.\ 2007, \apjl, 660, L1 
\bibitem[Donley et al.(2005)]{don05} Donley, J.~L., Rieke, G.~H., Rigby, J.~R., \& P{\'e}rez-Gonz{\'a}lez, P.~G.\ 2005, \apj, 634, 169 
\bibitem[Donley et al.(2007)]{don07} Donley, J.~L., Rieke, G.~H., P{\'e}rez-Gonz{\'a}lez, P.~G., Rigby, J.~R., \& Alonso-Herrero, A.\ 2007, \apj, 660, 167 
\bibitem[Elvis et al.(1994)]{elv94} Elvis, M., et al.\ 1994, \apjs, 95, 1 
\bibitem[Engelbracht et al.(2007)]{eng07} Engelbracht, C.~W., et al.\ 2007, \pasp, 119, 994
\bibitem[Fadda et al.(2002)]{fad02} Fadda, D., Flores, H., Hasinger, G., Franceschini, A., Altieri, B., Cesarsky, C.~J., Elbaz, D., \& Ferrando, P.\ 2002, \aap, 383, 838 
\bibitem[Fazio et al.(2004)]{faz04} Fazio, G.~G., et al.\ 2004, \apjs, 154, 10 
\bibitem[Fiore et al.(2007)]{fio07} Fiore, F., et al.\ 2007, ArXiv e-prints, 705, arXiv:0705.2864 
\bibitem[Georgakakis et al.(2008)]{geo08} Georgakakis, A., et al. \ 2008, MNRAS in press, ArXiv e-prints, 801, arXiv:0801:2160 
\bibitem[Gilli et al.(2007)]{gil07} Gilli, R., Comastri, A., \& Hasinger, G.\ 2007, \aap, 463, 79 
\bibitem[Hornschemeier et al.(2003)]{hor03} Hornschemeier, A.~E., et al.\ 2003, \aj, 126, 575 
\bibitem[Ivezi{\'c} et al.(2002)]{ive02} Ivezi{\'c}, {\v Z}., et al.\ 2002, \aj, 124, 2364 
\bibitem[Ivison et al.(2007)]{ivi07} Ivison, R.~J., et al.\ 2007, \apjl, 660, L77 
\bibitem[Jiang et al.(2007)]{jia07} Jiang, L., Fan, X., Ivezi{\'c}, {\v Z}., Richards, G.~T., Schneider, D.~P., Strauss, M.~A., \& Kelly, B.~C.\ 2007, \apj, 656, 680 
\bibitem[Kellermann et al.(1989)]{kel89} Kellermann, K.~I., Sramek, R., Schmidt, M., Shaffer, D.~B., \& Green, R.\ 1989, \aj, 98, 1195 
\bibitem[Kov{\'a}cs et al.(2006)]{kov06} Kov{\'a}cs, A., Chapman, S.~C., Dowell, C.~D., Blain, A.~W., Ivison, R.~J., Smail, I., \& Phillips, T.~G.\ 2006, \apj, 650, 592 
\bibitem[Lacy et al.(2004)]{lac04} Lacy, M., et al.\ 2004, \apjs, 154, 166 
\bibitem[Laird et al.(2006)]{lai06} Laird, E.~S., Nandra, K., Hobbs, A., \& Steidel, C.~C.\ 2006, \mnras, 373, 217 \bibitem[Maccacaro et al.(1988)]{mac88} Maccacaro, T., Gioia, I.~M., Wolter, A., Zamorani, G., \& Stocke, J.~T.\ 1988, \apj, 326, 680 
\bibitem[Mart{\'{\i}}nez-Sansigre et al.(2005)]{mar05} Mart{\'{\i}}nez-Sansigre, A., Rawlings, S., Lacy, M., Fadda, D., Marleau, F.~R., Simpson, C., Willott, C.~J., \& Jarvis, M.~J.\ 2005, \nat, 436, 666 
\bibitem[Mart{\'{\i}}nez-Sansigre et al.(2007)]{mar07} Mart{\'{\i}}nez-Sansigre, A., et al.\ 2007, \mnras, 379, L6 
\bibitem[Mukai(1993)]{muk93} Mukai, K.\ 1993, Legacy, 3, 21 
\bibitem[Mushotzky(2004)]{mus04} Mushotzky, R.\ 2004, Supermassive Black Holes in the Distant Universe, 308, 53 
\bibitem[Nandra et al.(1997)]{nan97} Nandra, K., George, I.~M., Mushotzky, R.~F., Turner, T.~J., \& Yaqoob, T.\ 1997, \apj, 477, 602 
\bibitem[Nandra et al. (2007)]{nan07} Nandra, K., 2008, in prep
\bibitem[Piccinotti et al.(1982)]{pic82} Piccinotti, G., Mushotzky, R.~F., Boldt, E.~A., Holt, S.~S., Marshall, F.~E., Serlemitsos, P.~J., \& Shafer, R.~A.\ 1982, \apj, 253, 485 
\bibitem[Ranalli et al.(2003)]{ran03} Ranalli, P., Comastri, A., \& Setti, G.\ 2003, \aap, 399, 39 
\bibitem[Rieke \& Lebofsky(1981)]{rie81} Rieke, G.~H., \& Lebofsky, M.~J.\ 1981, \apj, 250, 87 \bibitem[Rieke et al.(2004)]{rie04} Rieke, G.~H., et al.\ 2004, \apjs, 154, 25 
\bibitem[Risaliti et al.(1999)]{ris99} Risaliti, G., Maiolino, R., \& Salvati, M.\ 1999, \apj, 522, 157 
\bibitem[Seymour et al.(2007)]{sey07} Seymour, N., et al.\ 2007, \mnras, submitted
\bibitem[Stern et al.(2005)]{ste05} Stern, D., et al.\ 2005, \apj, 631, 163 
\bibitem[Szokoly et al.(2004)]{szo04} Szokoly, G.~P., et al.\ 2004, \apjs, 155, 271 
\bibitem[Treister et al.(2005)]{tre05} Treister, E., et al.\ 2005, \apj, 621, 104 
\bibitem[Ueda et al.(2003)]{ued03} Ueda, Y., Akiyama, M., Ohta, K., \& Miyaji, T.\ 2003, \apj, 598, 886 
\bibitem[Urry \& Padovani(1995)]{urr95} Urry, C.~M., \& Padovani, P.\ 1995, \pasp, 107, 803 
\bibitem[Vlahakis et al.(2007)]{vla07} Vlahakis, C., Eales, S., \&
  Dunne, L.\ 2007, \mnras, 379, 1042 
\bibitem[Weedman et al.(2006)]{wee06} Weedman, D.~W., Le Floc'h, E., Higdon, S.~J.~U., Higdon, J.~L., \& Houck, J.~R.\ 2006, \apj, 638, 613 
\bibitem[Werner et al.(2004)]{wer04} Werner, M.~W., et al.\ 2004, \apjs, 154, 1 
\bibitem[Willott et al.(2003)]{wil03} Willott, C.~J., Rawlings, S., Jarvis, M.~J., \& Blundell, K.~M.\ 2003, \mnras, 339, 173 
\bibitem[Worsley et al.(2005)]{wor05} Worsley, M.~A., et al.\ 2005, \mnras, 357, 1281 
\bibitem[Wu et al.(2005)]{wu05} Wu, H., Cao, C., Hao, C.-N., Liu, F.-S., Wang, J.-L., Xia, X.-Y., Deng, Z.-G., \& Young, C.~K.-S.\ 2005, \apjl, 632, L79 
\bibitem[Yun et al.(2001)]{yun01} Yun, M.~S., Reddy, N.~A., \& Condon, J.~J.\ 2001, \apj, 554, 803 
\end{thebibliography}

\clearpage

\begin{deluxetable}{ccccc}
\tabletypesize{\scriptsize}
\tablecaption{Comparison of Selection Techniques\label{tbl-1}}
\tablewidth{0pt}
\tablehead{
\colhead{} & \colhead{Number} & \colhead{X-ray} &
\colhead{IRAC Power-law} &  \colhead{Optically Identified} \\
\colhead{} & \colhead{in Sample} & \colhead{Detection} &
\colhead{Detection} & \colhead{AGN Spectra}
}
\startdata
Flux-Cut   & 28 & 4/26  & 5/26  & 1/2 \\
$q$-Selected & 47 & 15/37 & 6/28  & 2/6 \\
Total      & 72 & 19/61 & 10/53 & 3/8 \\
\enddata
\tablecomments{The first column gives the number of sources selected
  with the radio/mid-IR methods outlined in this paper.  The
  subsequent columns indicate the number of sources from
  our candidate list identified as AGNs via other selection
  techniques and the number of sources detected in the relevant parent
  sample.} 
\end{deluxetable}

\clearpage

\begin{figure}
\plotone{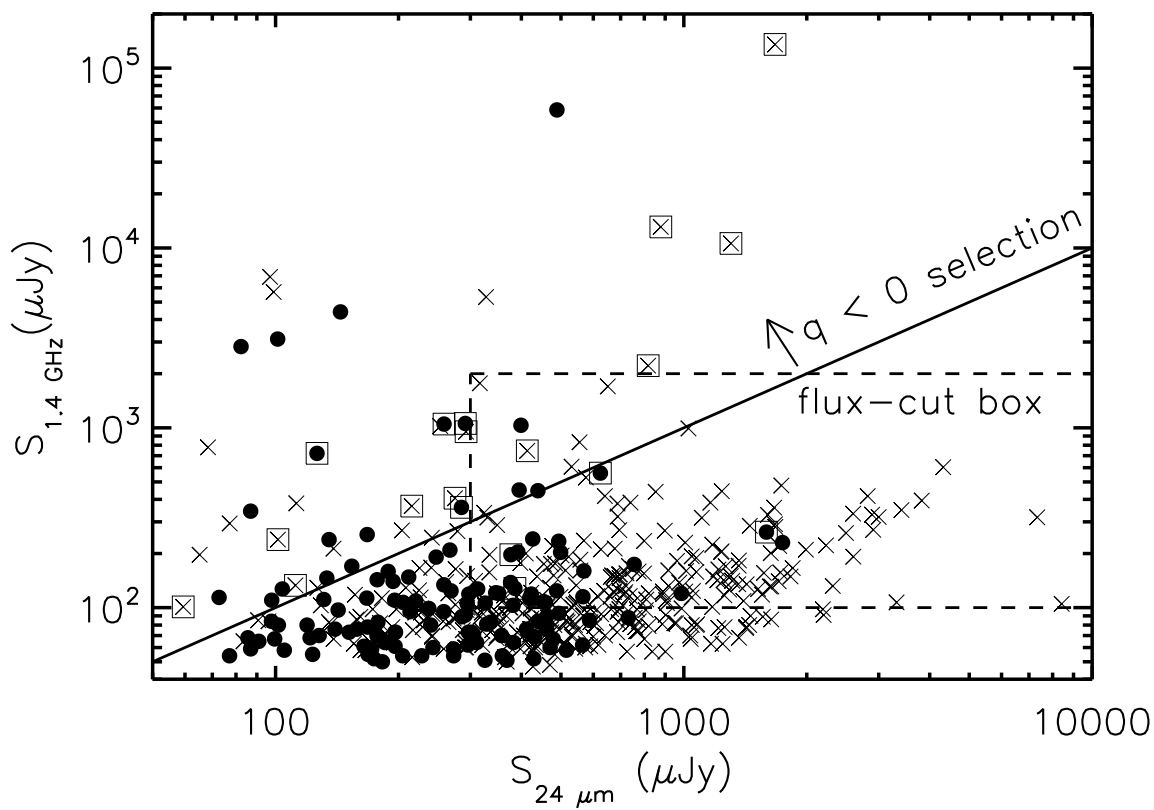}
\caption[f1.eps]{ Selection Criteria. All sources with both 24~\micron\
  and 1.4~GHz detections are plotted; dots represent sources with
  S$_{3.6~\mum} \le 45~\mu$Jy, all other sources are depicted by crosses.  
  AGN candidate sources with X-ray detections have been overplotted
  with squares. The flux-cut method selects all the
  dots within the dashed selection box; the $q$-selection method
  selects all sources above the solid line (dots and crosses).
  } 
\end{figure}

\begin{figure}
\plotone{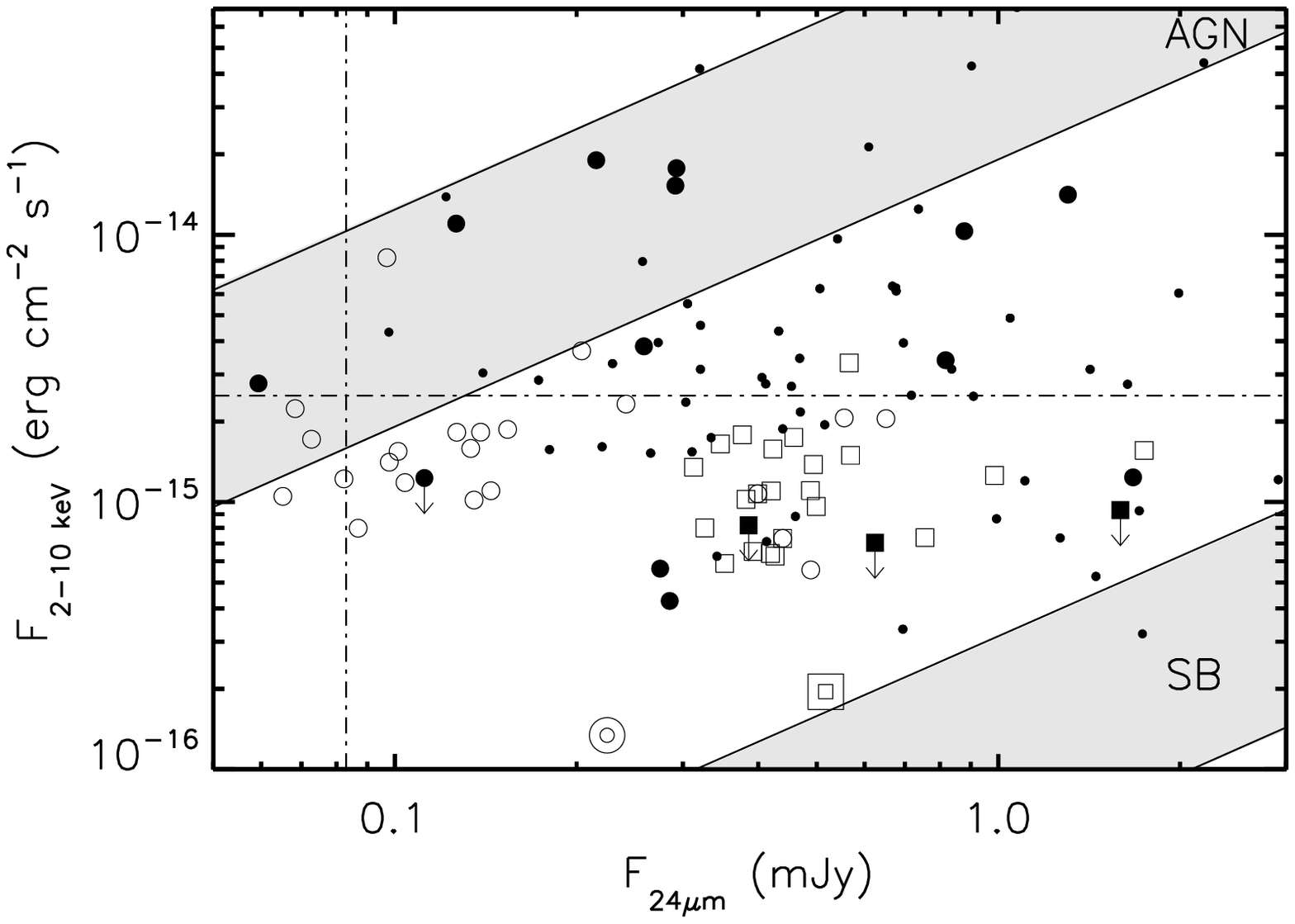}
\caption[f2.eps]{ 24~\micron\ emission vs.~the hard $2-10$ keV X-ray
fluxes for the sample of radio detected sources.  Squares/large
circles represent the flux-cut/$q$-selected sources.  All other hard
X-ray sources with 24~\micron\ detections in the field are depicted by
small dots.  Filled symbols denote sources with X-ray detections while
open symbols indicate hard X-ray upper limits for the non-detected
sources.  Filled symbols with arrows represent upper limits for
sources detected in the X-ray, but not in the hard band.  All open
symbols without arrows are X-ray upper limits; down-facing arrows were
not plotted for clarity.  The two stacked circles/squares represent
the hard band flux of the stacked signals and the mean 24~\micron\
emission of the stacked sources.  The horizontal line represents the
flux to which 80\% of the X-ray survey area is sensitive, and the
vertical line is the 80\% completeness limit for MIPS.
Extrapolated regions for local AGNs and starburst
galaxies are overplotted \citep{pic82, ran03, alo04}.  }
\end{figure}

\begin{figure}
\plotone{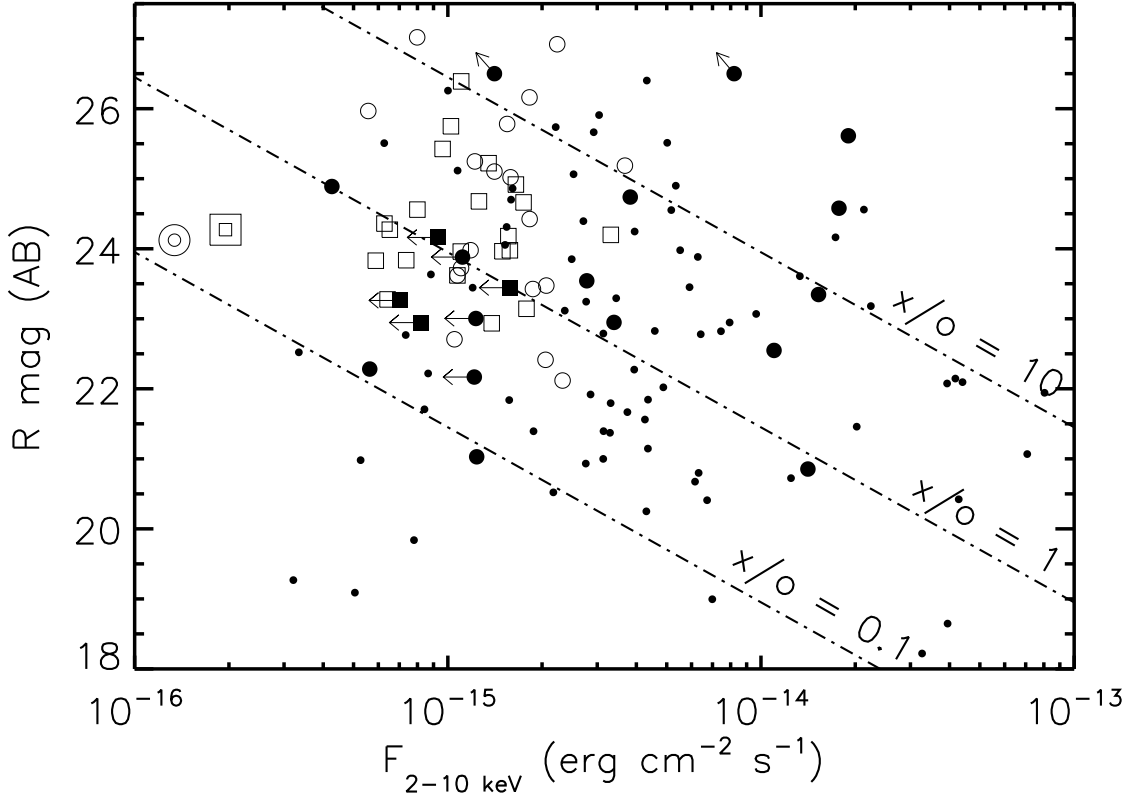}
\caption[f3.eps]{ Hard X-ray ($2-10$ keV) fluxes vs.~\emph{R} band
magnitudes.  Squares/large circles represent the flux-cut/$q$-selected
sources.  All other hard X-ray sources with $R$ band detections in the
field are depicted by small dots.  Filled symbols denote sources with
X-ray detections while open symbols indicate hard X-ray upper limits
for the non-detected sources.  Filled symbols with arrows represent
upper limits for sources detected in the X-ray, but not in the hard
band.  Symbols with arrows pointing upper left depict candidates
without firm detections in either hard X-rays or $R$ band.  All open
symbols without arrows are X-ray upper limits; left-facing arrows were
not plotted for clarity.  The two stacked circles/squares represent
the hard band flux of the stacked signals and the mean $R$ band
emission of the stacked sources.  Diagonal lines represent constant
value X/O ratios. AGNs are expected to lie above X/O $> 0.1$
\citep{hor03}}
\end{figure}

\end{document}